\documentstyle[twoside,fleqn,espcrc2]{article}
\def\nutau{ \nu_\tau }
\def\dm2{ \Delta m^2 }
\def\s22t{ \sin^22\theta }

\def\beq{ \begin{equation} }
\def\eeq{ \end{equation} }
\def\beqa{\begin{eqnarray} }
\def\eeqa{\end{eqnarray} }
\def\beqa*{ \begin{eqnarray*} }
\def\eeqa*{ \end{eqnarray*} }

\newcommand{\ttbs}{\char'134}
\newcommand{\AmS}{\protect\the\textfont2
    A\kern-.1667em\lower.5ex\hbox{M}\kern-.125emS}

\title{Theoretical Implications of the Combined
         Solar Neutrino Observations}
\author{{S.A. Bludman, N. Hata, D.C. Kennedy,
         P.G. Langacker}\\
   {\em Department of Physics,
   University of Pennsylvania, Philadelphia, PA 19104}\thanks
   { Supported in part by DOE Contract No. DOE-AC02-76-ERO-3071}}

\begin{document}
\begin{abstract}
Constraints on the core temperature of the Sun and on
neutrino-oscillation parameters are obtained by comparing
the combined Homestake, Kamiokande, SAGE and GALLEX
solar neutrino data with Standard Solar Models (SSM) and with
non-standard solar models parameterized by a
phenomenological central temperature ($T_c$).
If the Sun is 2\% cooler or 3\% warmer than predicted by
SSMs, the MSW parameters we determine are consistent with
different grand unified theories.

\end{abstract}

\maketitle

\section{Purely Astrophysical (Cooler Sun) Explanation for
the Observed Solar Neutrino Deficit is Most Unlikely}

Not only is a simultaneous fit of
the Homestake and Kamiokande data
incompatible with any temperature in the Sun at $>99.99\%$ C.L.,
but the central GALLEX value by itself
requires a 14\% cooler Sun and is already excluded at 65\% C.L.

\section{Matter-Amplified Electron-Neutrino Oscillations in the Sun}

If we assume this simplest particle physics
interpretation for a persistent
solar neutrino deficit, then the data and the SSM
constrain the MSW
parameters (neutrino masses, vacuum mixing angles) to lie in either of
two small regions: non-adiabatic
oscillations with
$\dm2 = (3 -  12) \ \mbox{meV}^2, \,
\s22t = (0.4 - 1.5)  \times 10^{-2}$,
or large mixing-angle oscillations with
$ \dm2 = (3 - 40) \  \mbox{meV}^2, \,
\s22t = 0.5 - 0.9 $. (Fig. 1)
The non-adiabatic solution gives the
considerably better fit.
Together with Homestake and Kamiokande,
the GALLEX experiment constrains
$\dm2 =(3-40)~\mbox{meV}^2$, which suggests
that low-energy $pp$ neutrinos are not oscillating appreciably.
The vacuum mixing angles fitted
are sensitive to the Sun's central temperature.

\section
{ MSW Fits With Non-Standard Core Temperatures}

If, indeed, $T_c=15.67\times 10^{6} $K, as obtained, for example,
in the latest Bahcall-Pinsonneault SSM with helium diffusion, then
both the small and large neutrino mixing angle fits in Fig. 1 differ
significantly from the corresponding (CKM) quark mixing angles.  If,
together with MSW oscillations, we
allow a non-standard core temperature, Fig. 2 shows that
the experiments themselves at 90\% C.L.
determine the core temperature:  relative to the theoretical value,
$T_c=1.03^{+0.03}_{-0.05}$ for the non-adiabatic fit,
$T_c=1.05^{+0.01}_{-0.07}$ for the large-angle fit, both consistent with
all SSMs.

Fig. 3 shows that for central temperatures within
7-9 (2-3)
times the theoretical uncertainty quoted by Bahcall and Ulrich
(Turck-Chi\`{e}ze {\it et al}),
the neutrino
parameters can fit in regions preferred by different grand
unified theories (GUTs):
A 2\% cooler Sun practically excludes the Cabibbo-angle solution
$\theta=\theta_{uc}$, but
extends $\dm2, \s22t$ to values $\theta=\theta_{ut}$
implied by the supersymmetric SO(10) GUT;
a $3-4\%$ warmer Sun practically excludes the $\theta=\theta_{ut}$
solution, but extends the
allowed parameter space to values $\theta=\theta_{uc}$
suggested by intermediate-scale
SO(10) GUT and for which the $\nutau$ may be cosmologically significant.
Superstring-inspired models are consistent with all parameter fits.

\section{Neutrino Spectral Distortion}

In the large-angle MSW solution, the
$\nu_e$ suppression is approximately energy-independent.
For the energies
to be detected at the Sudbury Neutrino Observatory or Super-Kamiokande,
the non-adiabatic solution suppresses
low-energy $\nu_e$ more than high-
energy neutrinos.  Consequently, such observations
can confirm the MSW
effect and, by distinguishing between the
two spectral shapes predicted in Fig. 4, can choose

\newpage
\noindent
between the non-adiabatic and large-angle solutions allowed by present
experiments.

\end{document}